# Adaptive AI decision interface for autonomous electronic material discovery

Yahao Dai[1,*], Henry Chan[2,✉], Aikaterini Vriza[2,*], Fredrick Kim[2], Yunfei Wang[3,4], Wei Liu[1], Naisong Shan[1], Jing Xu[2,5], Max Weires[1], Yukun Wu[2], Zhiqiang Cao[4], C. Suzanne Miller[2], Ralu Divan[2], Xiaodan Gu[4], Chenhui Zhu[3], Sihong Wang[1,2,✉], Jie Xu[1,2,✉]

[1] Pritzker School of Molecular Engineering, The University of Chicago, Chicago, IL, USA.

[2] Nanoscience and Technology Division, Argonne National Laboratory, Lemont, IL, USA.

[3] Advanced Light Source, Lawrence Berkeley National Laboratory, Berkeley, CA, USA.

[4] School of Polymer Science and Engineering, Center for Optoelectronic Materials and Devices, The University of Southern Mississippi, Hattiesburg, MS, USA.

[5] Present address: Department of Physics, University of Central Florida, Orlando, FL, USA.

[*] Those authors contribute equally.

[✉] e-mail: xuj@anl.gov; sihongwang@uchicago.edu; hchan@anl.gov.

## ABSTRACT

**AI-powered autonomous experimentation (AI/AE) can accelerate materials discovery but its effectiveness for electronic materials is hindered by data scarcity from lengthy and complex design-fabricate-test-analyze cycles[1]. Unlike experienced human scientists, even advanced AI algorithms in AI/AE lack the adaptability to make informative real-time decisions with limited datasets[2–5]. Here, we address this challenge by developing and implementing an AI decision interface on our AI/AE system. The central element of the interface is an AI advisor that performs real-time progress monitoring, data analysis, and interactive human-AI collaboration for actively adapting to experiments in different stages and types. We applied this platform to an emerging type of electronic materials—mixed ion-electron conducting polymers (MIECPs)—to engineer and study the relationships between multiscale morphology and properties. Using organic electrochemical transistors (OECT) as the testing-bed device for evaluating the mixed-conducting figure-of-merit—**

**the product of charge-carrier mobility and the volumetric capacitance ($\mu C^*$), our adaptive AI/AE platform achieved a 150% increase in $\mu C^*$ compared to the commonly used spin-coating method, reaching 1,275 F cm$^{-1}$ V$^{-1}$ s$^{-1}$ in just 64 autonomous experimental trials. A study of 10 statistically selected samples identifies two key structural factors for achieving higher volumetric capacitance: larger crystalline lamellar spacing and higher specific surface area, while also uncovering a new polymer polymorph in this material.**

## Main

Electronic materials make one of the most complicated, diverse, and useful classes of materials, which are designed based on the multi-level relationships between chemical structures, crystalline structures, electronic band structures, and properties spanning multiple aspects (semiconducting, conducting, optoelectronic, ion-conducting, dielectric, etc.)[6–10]. The rapid growth of global electronics demands significant advancement in electronic material innovation, yet, designing new materials with different targeted properties often requires years of extensive design, testing, and validation, which has been a bottleneck for the development and commercialization of new electronic technologies. To accelerate the research and development of electronic materials, recent advancements in AI-powered autonomous experimentation (AI/AE) have shown immense potential[11–18]. By integrating automated experimentation with data-driven planning, AI/AE can rapidly guide the next steps in an experimental campaign, expediting the hypothesis-discovery process[19–21]. However, the use of AI/AE in electronic materials discovery has been impeded by one major challenge – data scarcity[2], which results from the lengthy and complex design-fabricate-evaluate-optimize cycle in electronic materials, thereby limiting the effectiveness of predictive, data-driven design strategies.

To overcome this challenge, one possibility is to increase the data amount by increasing the capability of AI/AE systems. For electronic materials in particular, the systems need to integrate multiple automated workflows that connect material composition, processing, structure characterizations, and resulting property measurements. Among these, the challenging part is often the property measurements for electronic materials, many of which (e.g., charge carrier mobility) can only be accurately determined using device architectures (e.g., transistors) that replicate the conditions of their actual application in electronic systems. The integration of such device-level measurement capabilities into AI/AE systems requires automated fabrication processes for multilayer devices and sophisticated circuitry for device measurements[21]. This contrasts with ongoing efforts in the use of AI/AE on electronic materials, which primarily focus on property measurements of solutions or thin films, and thus could be limited to basic properties such as optical properties, crystallization behavior, and approximate charge carrier mobility[22–25]. Even though the development of more capable and parallel AI/AE systems could help to mitigate the data-scarcity challenge, the speed of data generation from such complicated and time-consuming cycles of design-fabricate-test-analyze may remain slow, given the typical resources available to most research groups. Therefore, we propose a pragmatic strategy that can radically reduce the reliance on big data in AI/AE systems to achieve the same performance.

Unlike experienced human scientists, even state-of-the-art AI algorithms used in AI/AE lack the adaptability to perform informative real-time decision-making based on small datasets[1,2,26]. Actually, human cognition excels in this regard[27]. As such, to solve the challenge of small datasets, we propose a strategy of fostering human-AI collaboration in AI/AE for electronic materials development. This strategy could also help mitigate concerns about the trustworthiness of AI[28,29]. Inspired by financial trading, known for data-driven decisions and dynamic strategy adaptation in high-frequency trading, we designed an AI advisor to quantitatively monitor the progress of AI/AE experiments and perform real-time data crunching tasks such as comparing the performance of AI algorithms. The AI advisor provides data-driven insights to enable adaptive changes in an AI/AE experimental workflow and refinements of parameter space by human scientists. We hypothesize that this platform can eventually lead to automated experimentation strategies that mimic human adaptability in making real-time, informed decisions based on small datasets, while simultaneously reducing cognitive biases such as anchoring and confirmation bias through mathematics and statistical data analysis.

In this work, we applied this innovation within our AI/AE laboratory, Polybot[30], to advance the development of an emerging class of electronic materials—mixed ion-electron conducting polymers (MIECPs)[31,32] (Fig. 1), which enable simultaneous transport and coupling of ions and electrons, essential for electrochemical devices in biosensing, neuromorphic computing, and energy storage. One primary device platform for the use of these materials is organic electrochemical transistors (OECTs)[33] that provide small voltage signal amplification and modulation. The key figure-of-merit for such functions of OECTs is transconductance ($G_m$), which is conjunctionally determined by two properties of MIECPs, i.e., the product of charge carrier mobility ($\mu$) and volumetric capacitance ($C^*$). In this regard, a significant amount of work has been dedicated to designing new chemical structures to enhance $\mu C^*$[34]. In addition to chemical structures, the wide variability in assembled polymer packing structures can significantly influence mixed ionic/electronic transport and their coupling within different length scales[35–37]. However, such morphological effects remain underexplored, particularly concerning its influence on $C^*$. As such, this presents a substantial opportunity to enhance mixed-conducting properties through engineering hierarchical structures.

By implementing a dynamic and adaptive decision-making strategy in Polybot and utilizing it to engineer the morphological structures of previously reported high-performance MIECP[38], we realized a 150% increase of $\mu C^*$ to 1,275 F cm$^{-1}$ V$^{-1}$ s$^{-1}$ compared to the commonly used spin-coating method, within just 64 autonomous experiment trials. The obtained data covered a broad $\mu C^*$ window, from 166 to 1,275 F cm$^{-1}$ V$^{-1}$ s$^{-1}$. Importantly, such a broad distribution of $\mu C^*$ enables the study of the relationship between hierarchical structures and mix-conducting properties for MIECPs. By using unbiased statistical data analytics to select 10 representative samples for further morphological characterizations, we reveal that two nanostructural features- larger crystalline lamellar lattice spacing and increased interfaces through nanofibrillar morphologies, substantially enhance $C^*$, which fills a major knowledge gap. Additionally, we also discovered the coexistence of two polymorphic structures of this MIECP for the first time.

## Automated workflow with a dynamic and adaptive decision strategy

To develop high-performance MIECP films for OECT, we hypothesized that the preferred morphology of MIECPs for achieving the optimal $\mu C^*$ relies on three key features: (1) planar chain conformation to extend π-electron delocalization along the polymer backbone, (2) sufficient $\pi$-$\pi$ stacking to enable efficient interchain electronic charge carrier hopping, (3) maximized interfacial areas for maximized coupling between electrons and ions. However, achieving control over the assembly of MIECPs into the hypothesized film morphology remains challenging, due to the very large parameter space of solution processing conditions (Supplementary Note 1).

To access such a large number of processing conditions while facilitating the formation of planar chain conformation, we design our AI/AE platform, Polybot, to perform solution-shearing deposition[39] of MIECP films on nanogrooved substrates, with the conditions informed by autonomous device characterizations on the films (Supplementary Figs. 1 and 2). This platform can be divided into two automated workflows (Fig. 1 and Supplementary Video 1). The first workflow focuses on the deposition of MIECP films and the subsequent OECT device fabrication, which includes a substrate/device storage station, solution preparing station, film coating station, annealing station, and an OECT fabrication station for the high-precision placement of electrolyte and Ag/AgCl gate electrode. The substrates for the film depositions incorporate nanogrooved designs of varying dimensions to promote polymer chain extension[40,41] and are pre-deposited with source/drain (S/D) electrodes and pre-loaded in the storage station (Supplementary Note 2 and Supplementary Fig. 3). This workflow varies four solution-shearing parameters that can affect polymer-chain assembly processes[42]: MIECP solution concentration, coating temperature, coating speed, and substrate nanogroove dimension. The second automated workflow carries out device characterizations and $\mu C^*$ calculation. It comprises a film thickness measurement station, a device probe station, and an electrometer station. The $\mu C^*$ is automatically extracted from the peak $G_m$ of the measured OECT transfer curve and measured thickness of the MIECP film:

$$\mu C^* = \frac{G_m}{\left(\frac{W}{L}\right) \cdot d \cdot (V_{Th} - V_{gs})}$$

Where $W$, $L$, $d$, $V_{th}$, and $V_{gs}$ stand for channel width, channel length, channel thickness, threshold voltage, and the gate voltage at peak $G_m$, respectively (Supplementary Note 3). Under each condition, the $\mu C^*$ is acquired from two separate samples that passed a two-sample t-test (Supplementary Note 4).

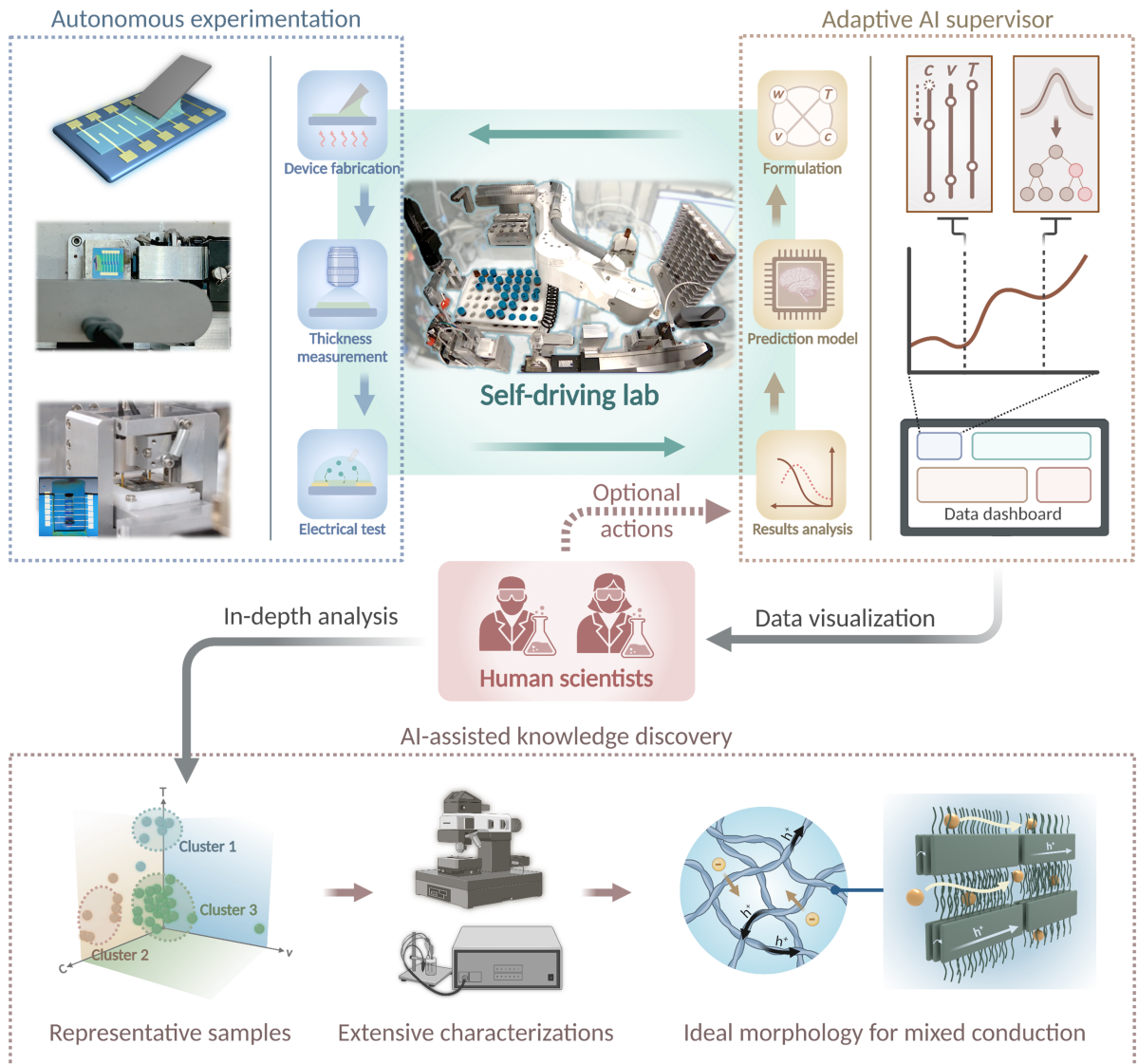

**Fig. 1. Utilizing AI/AE platform for expedited MIECP exploration.** The platform consists of an automated workflow for OECT fabrication and measurement and an AI advisor for decision intelligence by combining progress monitoring, generating actionable data insights, and adapting to new human inputs and insights over time. Following the autonomous exploration, the dataset is further used to generate representative samples in in-depth study for knowledge discovery.

The engineering of MIECPs' morphological structure for $\mu C^*$ enhancement through data-driven approaches requires substantial, high-quality datasets. However, obtaining those datasets in a short term remains a challenge, even with an AI/AE platform, due to the lengthy and time-consuming OECT fabrication-test cycles and rigid decision making approach. To address this challenge, we designed an adaptive and dynamic decision-making workflow (Fig. 2a) on Polybot, which seamlessly integrates an AI advisor for progress monitoring and real-time data analysis with interactive human-AI collaboration (Figs. 2b and 2c, Supplementary Video 2), ensuring an efficient and responsive system. The AI advisor, inspired by robo-advisors in the stock trading field, handles data crunching tasks and visually communicates any data trends, patterns, and anomalies, as well as ML model performance and feature importance (Supplementary Figs. 4-6) to scientists via a live-streaming platform (Supplementary Note 5). When a new data point is obtained, the AI advisor re-evaluates the experimental progress using several trend indicators (Supplementary Note 6) and retrains a set of ML prediction models (Supplementary Note 7 and Supplementary Table 1) to identify the most effective algorithm, and provides feature importance analysis for adaptive design space refinement (Fig. 2b). This information facilitates occasional human-AI interactions through a feedback mechanism, seamlessly integrating human intelligence into its critical decision-making process. This innovative workflow enhances the normal operations of a self-driving laboratory by aiding the perception and understanding of fundamental scientific principles from data while allowing adaptive changes informed by human knowledge and experience, efficiency, and adaptability.

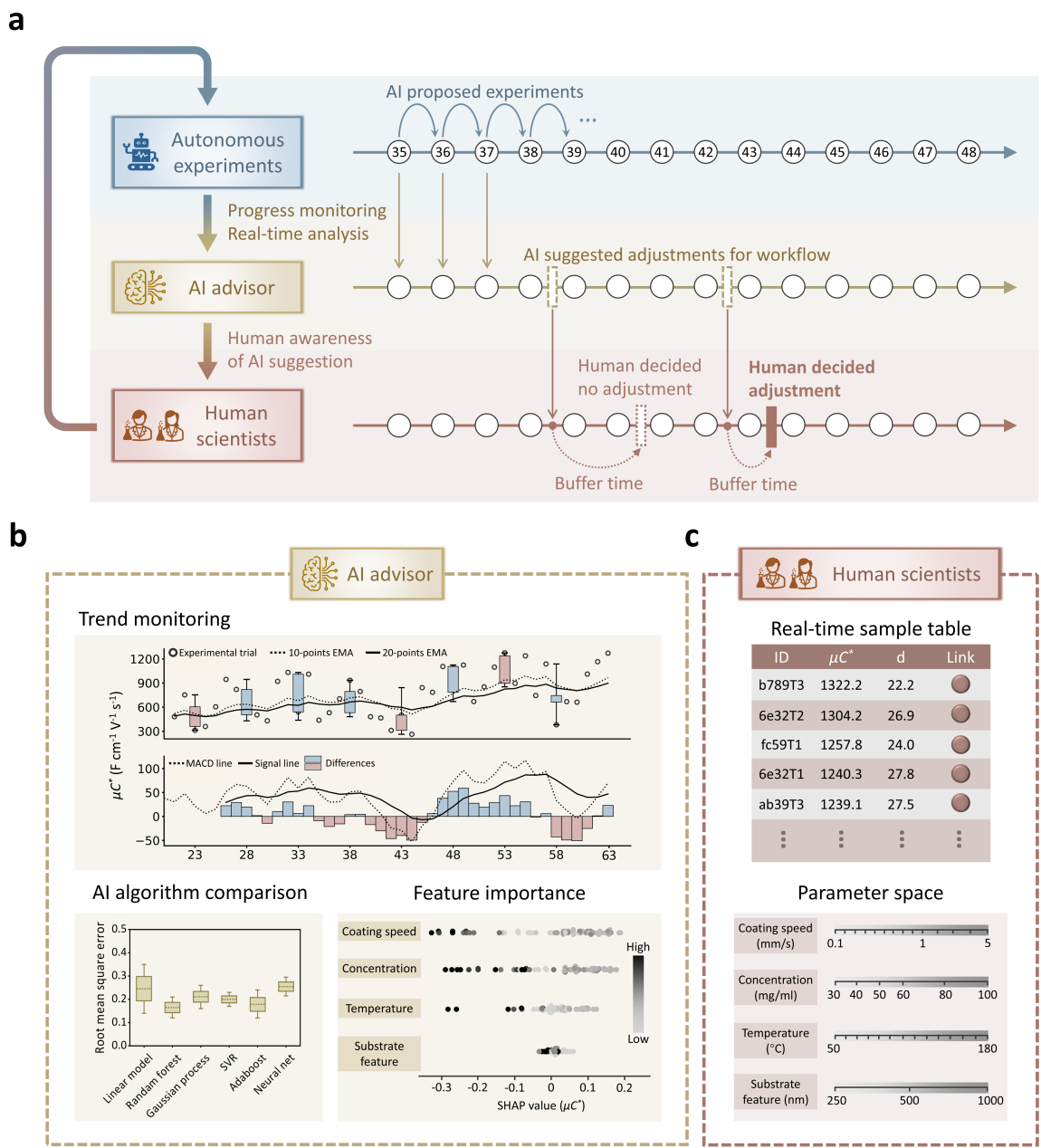

**Fig. 2. Real-time progress monitoring and analysis from AI advisor and its optional interaction with human scientists. a**, Schematic timeline of the interactions between AI and human scientists in the autonomous experimentation. Similar as typical self-driving lab, an AI algorithm proposes the sequence of the next experiments and autonomously runs in a closed-loop manner. In our case, additionally, an AI advisor automatically performs laborious data-crunching tasks while human scientists can simply observe and optionally modify the workflow based on the AI feedback. Although human-decided adjustments are informed by the AI advisor, they can be made anytime during the autonomous experiments with a certain

buffer time after AI-suggested timepoints. **b**, Exemplar tasks performed by the AI advisor include trend monitoring performed based on several indicators (Supplementary Note 7), prediction model comparison (Supplementary Table 1), and feature importance analysis (Supplementary Note 14). **c**, Human scientists can observe summary of the autonomous experimental data along with the information provided by the AI advisor, and decide workflow changes such as parameter space refinement based on AI suggestions.

## Autonomous experiments for $\mu C^*$ optimization through morphological engineering

Our objectives upon AI/AE platform-based MIECP exploration are two-fold: to establish a diverse set of morphological structures to understand the structural dependence of $\mu$ and $C^*$, and to figure out the optimal morphology for high-$\mu C^*$ MIECP film. We chose a high-performance MIECP, poly(2-(3,3′-bis(2-(2-(2-methoxyethoxy)ethoxy)ethoxy)-[2,2′-bithiophen]-5)yl thiophene) (p(g2T-T))[38] as the model MIECP (Fig. 3a and Supplementary Fig. 7). 1-chloronaphthalene was used as the solvent due to its superior solvation effect on the conjugated backbone[43], which increases the flexibility to modulate the polymer assemble pathway through processing condition adjustment. Using Polybot, the shearing-deposited films and the fabricated OECT devices (Fig. 3b) give ideal OECT transfer curves (Supplementary Note 8 and Supplementary Fig. 8) with high consistency from the four channels under the same film (Fig. 3c). Based on a robust, monotonic relationship established between the film thickness and color (Supplementary Note 9 and Supplementary Fig. 9), the Polybot system measures the film thickness through film color imaging (Fig. 3d, top) to further extract $\mu C^*$ (Fig. 3d, bottom). The solution shearing carried out by the Polybot under the same conditions give high repeatability (Supplementary Fig. 10), which is essential for the use of data in the AI algorithm.

In the autonomous study of p(g2T-T), we started with the following parameter ranges to effectively tune and optimize the assembled morphology: (1) coating speed — [0.1, 10] mm/s with an increment of 0.1 - 1 mm/s, (2) coating temperature — [70, 180] °C with an increment 10 °C, (3) solvent concentration — {30, 40, 50, 60, 80, 100} mg/ml, and (4) substrate trench width — {250, 500, 1000} nm. The rationales behind the choice of these ranges are discussed in Supplementary Note 10 and Supplementary Table 2. The combination of these initial experiment parameters gives a search space of 4,320 possible experiments. During our experiments, these parameter ranges were adjusted based on the interactive AI-human collaborative workflow previously described. In the chosen parameter space, we selected 21 sets of conditions using the Latin hypercube sampling method (Supplementary Note 11 and Supplementary Table

3) for the initial training of ML models. The most robust and least overfitting model, i.e., Gaussian Process Regression (GPR), is selected as the surrogate property prediction model to initialize the autonomous search of higher $\mu C^*$ using Bayesian Optimization (BO) algorithm (Supplementary Note 12 and Supplementary Fig. 11). The use of BO algorithm[44] effectively balances exploration, by targeting experimental conditions with high uncertainty to uncover new possibilities, and exploitation, by validating predicted results with low uncertainty to refine known outcomes, thereby maximizes information gathering with minimal number of experiments.

With the use of expected improvements (EI) as the acquisition function (Supplementary Note 8), GPR identified high-performing regions of the experimental parameter space, and led to gradually increased $\mu C^*$ in the first 12 experimental trials (Fig. 3e). However, after 12 experimental trials, the algorithm approached a point of diminishing returns where additional experiments led to little or no improvements in the achieved device performance (Fig. 3e). The systematic diminishing trend of $\mu C^*$ was identified by the AI advisor, which led to adaptive changes (ML algorithm and/or the experimental parameter space) in the experimental workflow made by human scientists (Supplementary Note 5). Specifically, the AI advisor compared the performance of ML property prediction models (Supplementary Fig. 4), which includes linear regression, non-linear models (support vector regression (SVR), neural network), ensemble models (random forest (RF), AdaBoost), and GPR. RF, a more data-intensive algorithm, was suggested as the best model based on the currently acquired experiments. At the same time, the AI supervision also provided insights crucial for the refinement of the experimental parameter space, based on the relationships and feature importance of the four experimental parameters identified through Shapley analysis (Supplementary Note 13, Supplementary Fig. 12, Supplementary Table 4).

After such adaptive changes, subsequent experimental trials re-established the steadily increasing trend of $\mu C^*$ (Fig. 3e). With just another ~20 experimental trials, Polybot achieved the highest $\mu C^*$ value of 1275 C $F^{-1}$ $cm^{-1}$ $s^{-1}$ since the beginning of the search, which is about 30% increase from the highest value obtained from GPR. This $\mu C^*$ value is also 50% higher than the highest result in the initial training group (Figs. 3e and 3f), and 150% higher than films prepared by standard spin-coating deposition (Supplementary Fig. 13). In contrast, if the workflow continued to use GPR ignoring the AI-suggested switching, the $\mu C^*$ continued to decline despite an additional number of experiments. This comparison demonstrates the effectiveness of incorporating adaptive adjustments within an autonomous experimental workflow. This ML model adjustment indicates that the prediction uncertainty of the GPR model is directly linked to unexplored regions of the parameter space, making it more effective for the parameter space sampling and maximizing information gathering during the initial stage. Once sufficient data had been collected, the ensemble learning strategy of the RF model proved more effective for providing robust property predictions and led to the observed drastic improvement in $\mu C^*$ value[45].

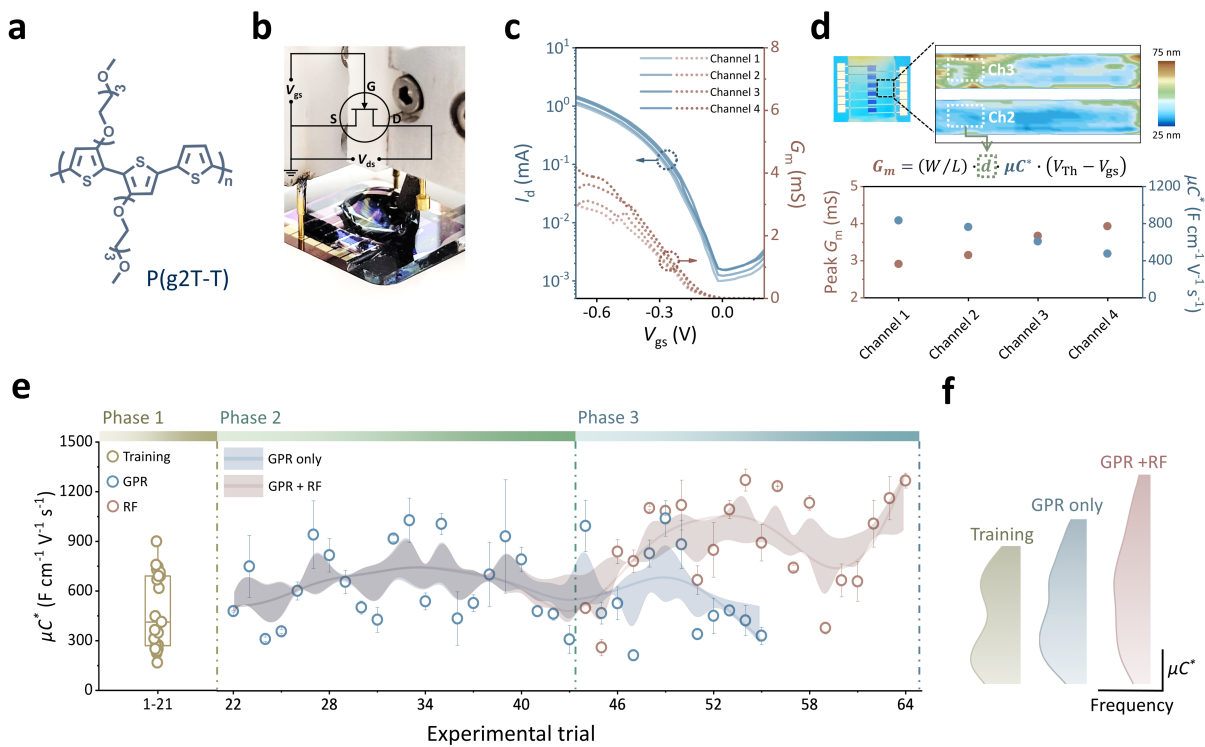

**Fig. 3. Autonomous MIECP exploration with AI/AE platform. a**, Molecular structure of p(g2T-T). **b**, Photography showing the OECT measurement with Polybot. **c**, Transfer curves of 4 channels within one chip. $V_{ds}$ = -0.8 V. **d**, Color-based thickness calculation (top) and measured peak $G_m$ and calculated $\mu C^*$ of 4 channels within one chip (bottom). **e**, Summary of the $\mu C^*$ trend in the multi-phase exploration: Phase 1. training phase, Phase 2. exploration phase with GPR as the prediction model, and Phase 3. exploration phase after space refinement, with either RF or GPR as the prediction model. Shaded regions behind the data points are overlays of trendlines with varying smoothness to help visually highlight the data trends. n = 6 independent measurements. **f**, $\mu C^*$ distributions of samples generated within the training phase or within exploration phases using different prediction models.

## Unbiased selection of samples with representative $\mu C^*$ and morphologies

The dataset generated from the MIECP films prepared under diverse processing conditions across a wide range of $\mu C^*$ also provides immense value in uncovering insights into their structure-property relationships.

Traditional approaches of characterizing all samples or relying on human selection become inefficient for large design space. In contrast, a statistics-guided selection could offer a more efficient approach by identifying a much smaller subset of "key" samples that effectively capture all the variational behaviors within the full sample space. To unbiasedly pick up the most representative samples, we combined two sampling methods, i.e., stratified sampling and cluster (K-means) sampling[46,47] (Fig. 4a, Supplementary Note 14). The purpose of stratified sampling is to segment the data distribution of the $\mu C^*$ into equally sized subgroups and select samples from each of these subgroups, while K-means sampling, a cluster-based technique, groups samples with similar experimental parameters and selects points from all the clusters for further analysis (Supplementary Fig. 14 and Supplementary Table 5). Using this set of methods, we identified 10 samples with the $\mu C^*$ values spanning the entire range and representing well-defined areas within the parameter space (Fig. 4b and Supplementary Table 6). We label these ten samples as S1-S10 (Supplementary Table 6, Supplementary Figs. 15 and 16). Together with samples prepared by spin-coated processes (labeled as S-C, Supplementary Fig. 13), these make into 11 sample conditions for in-depth characterizations.

$\mu$ and $C^*$ values were further characterized separately for these samples from each condition (Fig. 4c and Supplementary Fig. 17), using the constant gate current method (Supplementary Fig. 18) and the electrochemical impedance spectroscopy (EIS) measurement (Supplementary Fig. 19), respectively. The product of $\mu$ and $C^*$ agrees well with the $\mu C^*$ values calculated from the $G_m$ (Supplementary Fig. 20). Based on the distributions of $\mu$ and $C^*$, those samples cluster into three instinctive groups: Group 1: S1-S5, with high $\mu$ and high $C^*$; Group 2: S6, S7, with low $\mu$ but high $C^*$; and Group 3: S8-S10, and the S-C sample, with medium $\mu$ and low $C^*$ (Fig. 4c). We speculated that those major performance differences among these groups could be attributed to their significant structural differences.

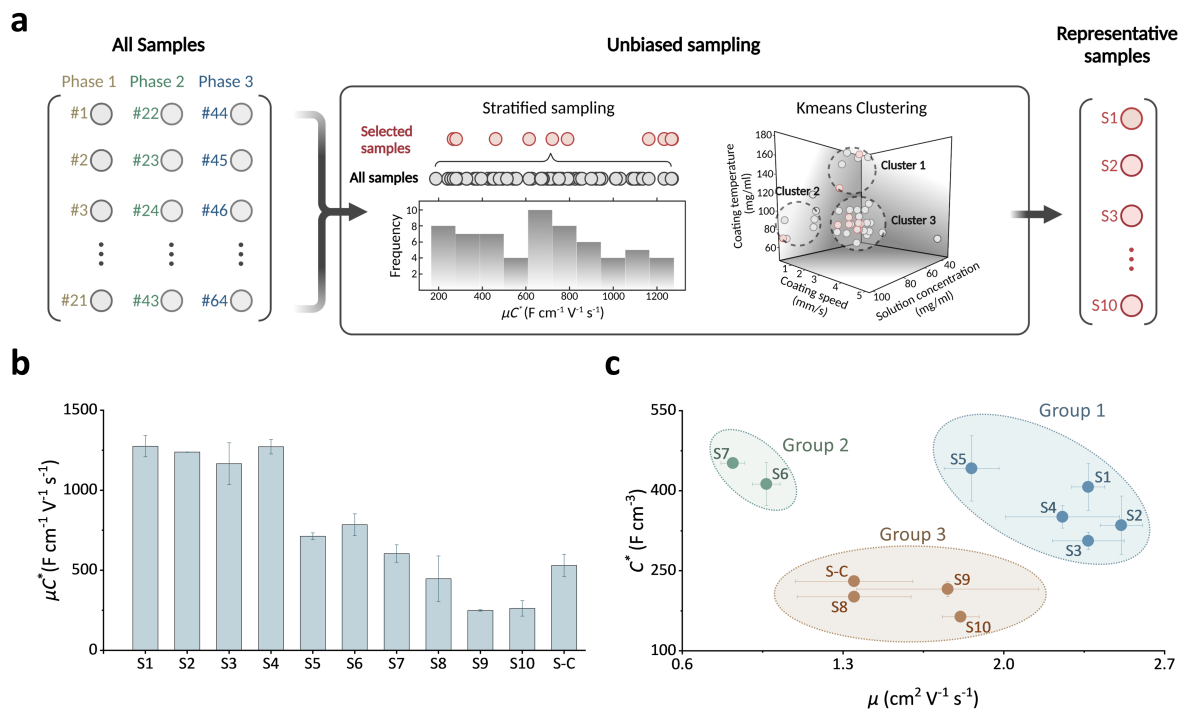

**Fig. 4. Unbiased selection of representative samples for in-depth characterizations**. **a**, A schematic showing the unbiased sampling of 10 representative samples (marked as red circles) from all experimental samples, which used the combined strategy of stratified sampling and K-means sampling. Stratified sampling segments the samples' $\mu C^*$ distribution into equally sized subgroups, ensuring balanced representation, while K-means sampling, a clustering technique, groups samples based on experimental parameters and selects representative samples from each cluster. **b**, Summarized $\mu C^*$ of representative samples. n = 6 independent measurements. **c**, Separately measured $\mu$ and $C^*$ of representative samples, which categorize samples into three distinct groups.

## Understanding structural dependence of $\mu C^*$

In-depth multi-scale morphological characterizations were conducted on these representative samples, encompassing analyses ranging from chain conformation and crystalline lattice structure to film morphology at the nano- to mesoscale. (Fig. 5a). From the study of polymer semiconductors in the past decades, it has been well-known that efficient electron transport (i.e., high $\mu$) should come from ordered structures with planar polymer chains and tight interchain packing[48,49]. However, less is known about morphological structure's influences on $C^*$, which relies on both ion transport and ion-electron coupling in MIECP films[50]. Here, we use three afore mentioned groups of samples to uncover these relationships. At the molecular level, we used Raman spectroscopy to characterize polymer chain conformational planarity. Higher planarity is reflected by a higher ratio of C=C (~ 1402 cm$^{-1}$)/C-C (~ 1439 cm$^{-1}$) stretching vibration peaks, as well as the redshift of both peaks[51]. As shown in Fig. 5b and Supplementary Fig. 21, the samples in Group 1 overall exhibit a more planar backbone conformation compared to those in the other two groups. This structural factor is consistent with the much higher $\mu$ observed in Group 1(Fig. 5c). However, no clear correlation with $C^*$ has been established based solely on the characterization of chain conformation.

For the crystalline morphology at the nanoscale, grazing incidence X-ray diffraction (GIXD) was used to characterize the lattice structures. From the diffraction patterns shown in Fig. 5d and Supplementary Fig. 22, the samples in Group 1 and Group 2 exhibit the same type of crystalline structure with similar lattice spacings. Differently, all the samples in Group 3 display a crystalline structure consisting of two distinct polymorphs, which, to the best of our knowledge, presents the first case of revealing two polymorphs within the single MIECP system. Compared to the crystalline structure of Group 1 and 2, this new crystalline structure found in Group 3 has a slightly larger π-π stacking distance of 3.55 Å (vs. 3.51 Å) (Fig. 5e and Supplementary Fig. 23), and two smaller lamellar distances of 14.2 Å and 17.5 Å (Fig. 5f and Supplementary Fig. 23). The distinct crystalline structures observed in Group 3, compared to Groups 1 and 2, clearly demonstrate that larger lamellar spacing contributes to increased $C^*$ values. (Fig. 5g and Supplementary Fig. 24).

At the nano- to mesoscale, we used AFM to characterize the microstructural morphology in the thin films. As shown in Fig. 5h and Supplementary Fig. 25, while nanofibrillar structures can be found in all the samples, the samples in Groups 1 & 2 have smaller fiber diameters compared to those in Group 3 (Fig. 5i). As such, the total electrolyte interface aspect ratios of the nanofibrous structures in Groups 1 & 2 should be much higher than those in Group 3. With Group 1 and 2 giving much higher $C^*$ values than Group 3 and faster ion penetration, as evidenced by the electrochemical impedance spectra, this validates that a larger electrolyte interface can facilitate ion-electron coupling and ion insertion within the MIECPs (Figs. 5j, 5k, and Supplementary Fig. 26).

In summary, we have created an AI decision interface that is seamlessly integrated into AI/AE, enabling a dynamical and adaptive decision workflow for electronic materials discovery, effectively facilitating AI-guided study using small datasets. By applying this methodology to MIECPs, we showed the high efficiency of this adaptive AI interface in optimizing the morphological structure of MIECP films and achieving greatly enhanced mixed-conducting properties just within a small number of (i.e., 64) experimental trials. In addition, these relatively small number of experimental results cover a large structure and property variation space, even including a previously unknown polymorph. This enabled us to uncover new insights into how morphological structures, including chain conformation, crystalline structures, and microscale film morphology, influence mixed-conducting properties, complementing existing knowledge on the role of chemical structures. Such new knowledge can serve as the basis for the morphological design and engineering of MIECPs toward higher performance. Overall, we anticipate that the adaptive AI decision-making concept can be broadly applied to AI/AE for other classes of electronic materials, offering similar benefits in achieving high efficiency for property enhancement and uncovering previously unknown fundamental relationships using limited datasets.

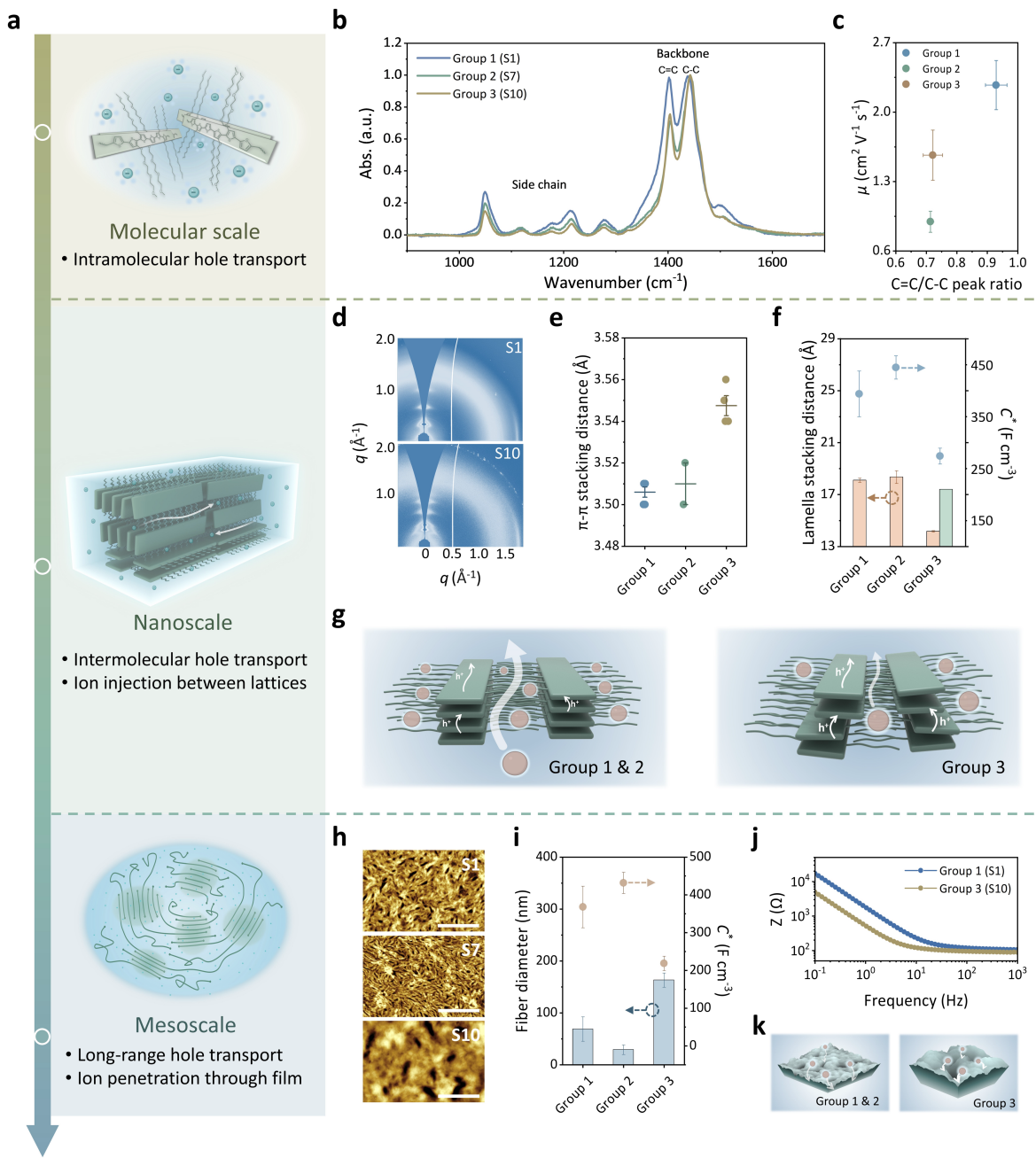

**Fig. 5. In-depth structural characterizations of the representative samples.** **a**, A schematic showing the ionic and electronic transport happening in different length scales. **b**, Representative Raman spectra of MIECP films from three groups. **c**, Positive correlations between $\mu$ and segmental ordering quantified by relative ratio between C=C and C-C stretching vibration peaks. **d**, Representative 2D GIXD patterns of MIECP films with single or mixed polymorphs. **e**, Summary of π-π stacking distance of MIECP films from different groups, which shows a negative correlation with $C^*$. **f**, Summary of lamella stacking distance of MIECP films from different groups, which shows a positive correlation with $C^*$. **g**, Schematics showing the critical role of intermolecular stacking distance for accommodating more ion injection into the crystalline domain, resulting in higher $C^*$. **h**, Representative nano-fibrillar morphology of MIECP films from three groups. Scale bars, 500 nm. **i**, Summary of averaged fiber diameters from each group of MIECP films. The group of samples showing lower diameters allows more efficient ion penetration, thus giving higher capacitance. **j**, Measurements of ion transport speed, which is characterized by the highest frequency the impedance response to. **k**, Schematics showing the influence of surface roughness on the ion penetration kinetics.

## Methods

**Materials**

All the chemicals in this work were used from purchase without further purification. 2,5-bis(trimethylstannyl)thiophene, bromobenzene, chlorobenzene, trimethyl(phenyl)tin, chloroform, hexane, ethyl acetate, methanol, tris(dibenzylideneacetone)dipalladium, tri(o-tolyl)phosphine, sodium chloride, 1-chloronaphthalene were purchased from Sigma-Aldrich. 5,5'-dibromo-3,3'-bis(2-(2-(2-methoxyethoxy)ethoxy)ethoxy)-2,2'-bithiophene was purchased from SunaTech Inc. Polydimethylsiloxane (Sylgard 184) was purchased from Ellsworth Adhesives.

**Synthesis of p(g2T-T)**

The polymerization of p(g2T-T) was modified according to the reported method[38], with the Biotage Initiator[+] microwave reactor. Briefly speaking, in a clean 5.0 mL microwave vial, 63.8 mg of 2,5-bis(trimethylstannyl)thiophene (155.7 µmol) and 100.0 mg of 5,5'-dibromo-3,3'-bisalkoxy-2,2'-bithiophene (155.7 µmol) were dissolved in 2.0 mL of anhydrous degassed chlorobenzene. $Pd_2(dba)_3$ (2.48 mg, 2.71 µmol) and $P(o-tol)_3$ (3.76 mg, 12.3 µmol) were added and the vial was sealed under nitrogen. The vial was subjected to heating at 80 °C for 10 min. After polymerization, the vial was cooled, 40 µL of trimethyl(phenyl)tin was added and the contents were subjected to microwave heating at 80 °C for 5 min. Finally, 100 µL of 2-bromobenzene was added and the reaction was subjected to microwave heating at 80 °C for 5 min. Then the reaction mixture was cooled to room temperature and precipitated in methanol. Formed blue solids were filtered by a filtering paper, and Soxhlet extraction was carried out with hexane and ethyl acetate for 12 h at each step. The polymer was then dissolved in hot chloroform. Finally, the polymer chloroform solution was concentrated and dried under a high vacuum. The molecular weight of p(g2T-T) was measured by Wyatt /Shimadzu size exclusion chromatography in the THF phase using polystyrene standards (Supplementary Fig. 7).

**Manual OECT fabrication and measurement**

To fabricate OECT devices, 10 nm Cr and 50 nm Au were deposited on nanogrooved $SiO_2$ substrates via shadow masks as the drain/source electrodes. The channel length and width were fixed as 1 mm and 2 mm, respectively. Then, neat p(g2T-T) films were deposited on top using either blade coating or spin coating method. For blade coating, the coating process was automatically conducted by Polybot with recipes of selected samples. For spin-coating, the films were coated using the p(g2T-T) with the concentration of 30 mg/ml in 1-chloronaphthalene at 3000 rpm. After film deposition, the films were quickly annealed at 150 °C under nitrogen for 5 min to fully dry film. Then the excessive semiconducting films outside the channel area were carefully removed using a blade under an optical microscope to isolate the device for measurement. For OECT measurement, poly(dimethylsiloxane) (PDMS) strips (with 10:1 base/crosslinker ratio) were placed on both sides of the channel to confine the electrolyte. Then, aqueous NaCl solution of 100 mM was carefully dropped on the channel. During the test, an Ag/AgCl pellet (Warner instruments, E210) was immersed in the electrolyte as the gate electrode. The device measurement was carried out using a Keithley 4200 semiconductor system.

**Mobility measurement based on OECT configuration**

The $\mu$ of OMIEC films under OECT operation was obtained by measuring the hole transit time ($\tau_h$) as reported[52]. Briefly speaking, different constant gate currents ($I_g$) have been applied to the device under constant drain bias ($V_{ds}$ = -0.1 V). Then the transient slope is plotted versus $I_g$ to calculate the $\tau_h$. The $\mu$ is calculated by:

$$\mu = \frac{L^2}{\tau_h \times V_{ds}}$$

where the $L$ is the channel length.

**Electrochemical impedance spectroscopy**

The electrochemical impedance spectra were collected using a PalmSens electrochemical workstation. The p(g2T-T) thin films were deposited on the planar Au electrode as a working electrode as aforementioned. The Au area uncovered by OMIEC film was then manually cut. Besides, Ag/AgCl was used as the reference electrode and a Pt plate was used as the counter electrode. EIS measurements were performed in 100 mM aqueous NaCl solution over a frequency range of 1 kHz to 0.1 Hz with an AC amplitude of 10 mV, and a DC offset of 0.2 V. The analysis of EIS data was carried out using Multitrace 4.2 software. The capacitance was extracted by fitting the EIS results with a simplified Randle model[53], and then normalized by film dimensions. Each $C^*$ was averaged from two samples.

**Grazing incidence X-ray diffraction (GIXD) measurement**

GIXD measurements were conducted on beamline 7.3.3 at the Advanced Light Source in Berkeley Lawrence National Laboratory. Data were collected under a Helium environment with an incident beam energy of 10 keV and an incidence angle of 0.15°. Thin-film samples were prepared on Si substrates. The scattering signals were collected by a Pilatus 2M detector and processed using Igor 8 software combined with the Nika package and WAXSTools.

**Other characterizations**

The thickness measurement was performed using either Tencor P-7 Stylus Profiler or Bruker Multimode 8 AFM. The AFM images of OMIEC films were acquired using Bruker Multimode 8 AFM under the tapping mode. The Raman spectra of the p(g2T-T) films were measured using a RENISHAW inVia confocal Raman microscop with the excitation beamline He−Ne laser of wavelength 632 nm and 0.1 mW power. The dispersion gratings were 600 and 1800 grooves/mm. The spectra were collected with a 100× objective.

# Methods References

# Data availability

All source data are provided within this paper and supplementary information. All relevant data are provided at the GitHub repository: https://github.com/polybot-nexus/robo_advisor_dashboard.

# Code availability

All the relevant code associated with this work including the AI-advisor platform, model training, evaluation, and data processing are publicly available via GitHub at https://github.com/polybot-nexus/robo_advisor_dashboard. An online version of the dashboard can be found at https://robo-advisor-dashboard.onrender.com.

# Acknowledgments

This work is supported by the U.S. Department of Energy, Office of Basic Energy Sciences, under Contract No. DE-AC02-06CH11357 (AI-guided robotic platform development) and the U.S. Department of Energy, Office of Science, Materials Sciences and Engineering Division (material design and characterization). Work performed at the Center for Nanoscale Materials, a U.S. Department of Energy Office of Science User Facility. This research used beamline 7.3.3 of the Advanced Light Source, which is a DOE Office of Science User Facility under contract no. DE-AC02-05CH11231. F.K. was supported in part by the Laboratory Directed Research and Development (LDRD) program at Argonne National Laboratory. Y.W. was supported in part by an ALS Doctoral Fellowship in Residence. X.G., Y. W., and Z. C. thank Office of Naval Research (ONR) under contract number N00014-23-1-2063 for providing funding to enable the X-ray based morphology characterization performed in this work. We thank David A. Czaplewski for the help with the fabrication of nanogrooved substrates.

# Author Contributions

Jie X., H.C., and S.W. supervised this work. Y.D., Jie X., H.C., A.V., and S.W. designed the experiments. H.C., Y.D., and Jie X. set up the hardware for automated OECT fabricating and testing. H.C. and A.V.

designed the code for automated workflow. A.V. and H.C. designed the AI advisor interface. H.C. and Y.D. set up the protocol for imaging-based thickness measurement. Y.D. and Yuk.W. synthesized and purified the polymers. N.S. carried out the size exclusion chromatography. F.K., Jin.X., C.S.M. and R.D. fabricated the nanogrooved substrates. Y.D. and M.W. determined the geometry and testing conditions for mitigating the short channel effect. A.V. and H.C. designed the protocol for the unbiased selection of representative samples. Y.D. fabricated the OECT and carried out mobility measurements. Y.D. performed the EIS measurements. Y.D., W.L., and N.S. performed the Raman spectroscopy measurement. Yun.W., C.Z., Z.C., and X.G. performed the GIXD measurements. Y.D. performed the AFM measurement. Y.D., H.C., A.V., S.W., and Jie X. wrote the paper. All authors reviewed and commented on the manuscript.

## Competing interests

The authors declare no competing interests.